# First principles investigation on anomalous lattice shrinkage of W alloyed rock salt GeTe

Guanjie Wang[a,b], Jian Zhou[a,b], Zhimei Sun[a,b,*]
[a]School of Materials Science and Engineering, Beihang University, Beijing 100191, China.
[b]Center for Integrated Computational Materials Engineering, International Research Institute for Multidisciplinary Science, Beihang University, Beijing 100191, China.

**ABSTRACT**

According to Vegard's law, larger radius atoms substitute for smaller atoms in a solid solution would enlarge the lattice parameters. However, by first-principles calculations, we have observed unusual lattice shrinkage when W replaces Ge in rock salt GeTe. We attribute this anomalous contract to the larger electronegativity difference between W and Te than between Ge and Te, which results in shorter W-Te bonds and pronounced local distortion around W dopants. The present work would provide new insight into the lattice parameter determination and a deeper understanding of the structural properties of ternary solid solutions.

## 1. Introduction

Intriguing narrow bandgap semiconducting chalcogenide GeTe have gained extensive attention as phase-change data storage media [1, 2]. Doping is a powerful way to improve its properties [3-5]. Upon doping, the change of lattice spacing will affect its elastic constants, thermal expansion and electrical conductivity. Thus, determining the valid lattice parameter of doped GeTe is of great importance. Vegard's law [6] has enjoyed long-lasting popularity by providing an estimation that the lattice parameters vary linearly with the solute size and concentration at a constant temperature when two constituents form a continuous solid solution, in which the solute atoms or ions distribute randomly among the solvent matrix. However, we have observed an abnormal shrinkage of GeTe when more massive W substitutes smaller Ge, which is in contrary to lattice expansion as predicted by Vegard's law. The origin of such counterintuitive observation remains ambiguous.

Several similar cases have been reported previously. For example, smaller Ge substitute larger Ni renders lattice dilation [7]. Fe-Ni binary alloy [8] also shows the opposite volume change, which could be explained by a numerical model based on interatomic spacings. However, this model requires that the interatomic spacings must be independent of other atoms in the neighborhood, which could be hardly fulfilled in ternary systems like W doped GeTe. In Ag-Au alloys [9], larger Ag substitute smaller Au decreases the lattice parameter due to electronic interactions between the outer electron shells of Ag and Au. This proposition provides us valuable clues that the radius of a solute atom would change by the transfer of electrons between high-energy electron shells of the solvent and the solute atoms.

Vegard's law could be violated when factors on the electronic level show a more significant impact on the lattice parameters. In the present work, we have investigated the structure of W alloyed GeTe from the perspective of chemical bonding on the electronic scale through first-principles calculations. We attribute the anomalous shrinkage to the electronegativity effects between W and Te, which render shorter W-Te bonds and denser local structures around W.

## 2. Computational methods

The *ab initio* calculations were performed within the framework of density functional theory (DFT) using projector augmented wave (PAW) pseudopotential method as implemented in the Vienna *ab initio* simulation package (VASP) [10]. The generalized gradient approximation of Perdew-Burke-Ernzerhof (PBE) and a cutoff of 300 eV were applied for the structural relaxation and self-consistent calculations [11]. Strongly constrained and appropriately normed meta-generalized gradient approximation (SCAN) has been reported to achieve remarkable accuracy for lattice constants [12]. Thus, it was applied to confirm the anomalous volume shrinkage obtained by PBE. Gaussian smearing with a smearing width of 0.02 eV was used to determine the partial occupancies. The Gamma centered *k*-mesh was set to $4 \times 4 \times 4$ for relaxation and *ab initio* self-consistent calculations.

The initial configuration is a rock salt 2 × 2 × 2 supercell of $W_mGe_{32-m}Te_{32}$ (m = 0, 1, 2, 3) containing 64 atoms (WGeTe), where W and Ge occupy one sublattice while Te ions occupy the other sublattice. WGeTe models with different configurations were relaxed using the conjugate-gradient method before performing *ab initio* self-consistent calculations. As full relaxations for WGeTe would induce structural collapse [13], we force the models to retain their original rock salt symmetry by implementing symmetry preserved relaxations in terms of cell volume and ionic positions, respectively. The relaxations were terminated when the energy difference between iterations was less than $1.0 \times 10^{-6}$ eV. According to our calculations for $W_mGe_{32-m}Te_{32}$ (m = 2, 3), the configurations that W dopants locate near to each other would result in lower total energy than randomly distributed ones. Therefore, only energetically favorable configurations are shown in the article.

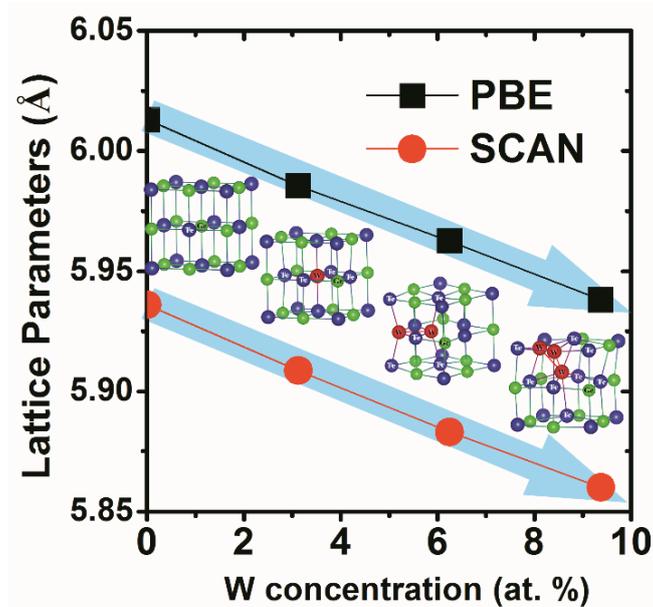

Fig. 1 Calculated lattice parameters of W-GeTe alloys. Results calculated by PBE were shown in black squares, and SCAN results were depicted by red circles, while the lines and arrows are guides for eyes. Insets are local structures around central Ge or W for $Ge_{32}Te_{32}$, $W_1Ge_{31}Te_{32}$, $W_2Ge_{30}Te_{32}$, $W_3Ge_{29}Te_{32}$

## 3. Results and Discussion

As shown in , the lattice parameter of W-GeTe decreases with increasing W concentration calculated by both PBE and SCAN. It can be seen that the lattice parameter of pristine GeTe calculated by PBE and SCAN is 6.013 Å and 5.936 Å, respectively. They are both in good agreement with the experimental data measured at 773 K (6.024 Å) [14]. The observation that the lattice parameter and cell volume decrease with the increasing concentration of larger radius dopants is in direct contrary to Vegard's law. Meanwhile, the average bond length ~ 2.747 Å for W-Te bonds is shorter than Ge-Te bonds (~ 2.951 Å) in $W_1Ge_{31}Te_{32}$, as shown in the inset of . Hence, W would not only induce cell volume shrinkage but render compact local structures as well. Furthermore, W-Te and W-W bonds distort the rock salt structure by introducing non-orthogonal bonds. The local distortion is so severe that the rock salt lattice could hardly be preserved when W concentration reaches 9.375 at.% ($W_3Ge_{29}Te_{32}$) as shown in the inset of . Therefore, we abandon models with W concentration higher than $W_3Ge_{29}Te_{32}$ in our symmetry preserved relaxations. The short and distorted W-Te bonds would be the origin of the peculiar volume shrinkage.

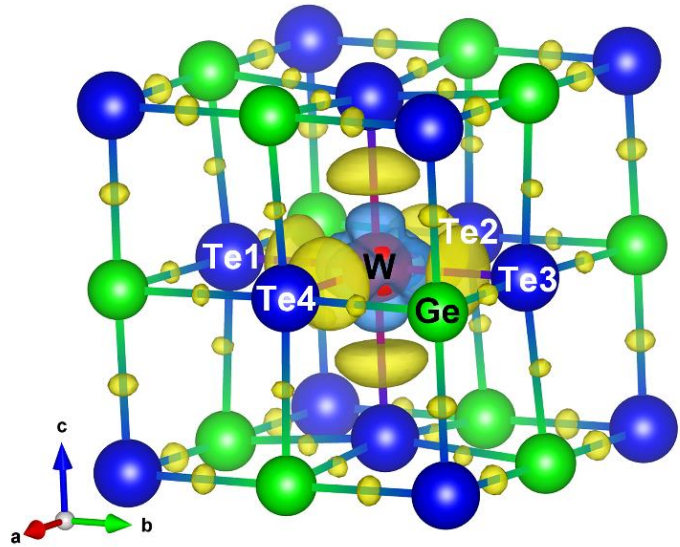

Fig. 2 Charge density difference (CDD) around W in $W_1Ge_{31}Te_{32}$. The yellow (blue) surfaces indicate 0.0055 e/bohr$^3$ increase (decrease) of charge density after bonds formation

As shown in , W is surrounded by blue surfaces, indicating that W loses its outer shell electrons (-0.0055 e/bohr$^3$) after bonding with six neighboring Te. On the other hand, yellow surfaces (+0.0055 e/bohr$^3$) are observed between W-Te, suggesting that several electrons of W transfer to Te, resulting in W-Te bonds. The not fully occupied 5$p$ orbitals of Te play a key role in accommodating the W outer shell electrons. Moreover, the yellow surfaces around W-Te are much larger than Ge-Te, indicating that more electrons are involved in W-Te bonding. This character results in strong W-Te interactions and hence short W-Te bonds. Therefore, we propose that W loses its outer shell electrons, which are accommodated in Te $5p$ orbitals when forming strong chemical interactions with Te. Consequently, the compact local structure around W and the abnormal cell volume shrinkage are observed.

Table 1 The average values of Bader charge transfer in WGeTe.

| System | Bader charge transfer/e | | |
| --- | --- | --- | --- |
|  | W | Ge | Te |
| GeTe | … | 0.451 | -0.451 |
| $W_1Ge_{31}Te_{32}$ | 0.616 | 0.425 | -0.431 |
| $W_2Ge_{30}Te_{32}$ | 0.505 | 0.394 | -0.401 |
| $W_3Ge_{29}Te_{32}$ | 0.505 | 0.350 | -0.364 |

Bader charge transfers can estimate electron transfer during bonds formation[15]. As shown in Table 1, W and Ge lose their electrons while Te obtains them. It is noteworthy that W loses 0.616 e in $W_1Ge_{31}Te_{32}$ when it is coordinated by six Te ions. On the other hand, the Bader charge transfer of W decreases when the first coordination shell includes another W (0.505 in $W_2Ge_{30}Te_{32}$ and $W_3Ge_{29}Te_{32}$). However, W still loses more electrons than Ge, whose Bader charge transfers are among 0.350 – 0.425 e. This might be attributed to the lower electron affinity of W nucleus to its outer shell electrons, namely the 2 electrons in the $6s$ orbital and the 4 electrons in the $5d$ orbital. Bader charge transfer data have demonstrated the former proposition that W loses outer shell electrons and induces short and strong W-Te bonds, which shrink the local structure as well as decrease cell volume of WGeTe.

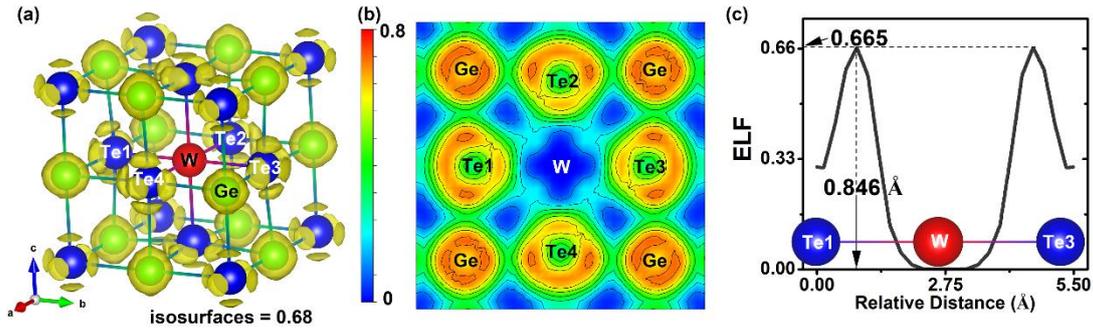

Fig. 3 Electron localization function (ELF) around W in $W_1Ge_{31}Te_{32}$. (a) Three-dimensional display of ELF, in which the yellow surfaces indicate ELF values higher than 0.68. (b) A two-dimensional slice of the (001) plane across W with ELF varies from 0 to 0.8, and the contour line interval was set to 0.1. (c) ELF line profile across Te1-W-Te3, where the relative distance is referred to Te1.

Further analysis of the electron localization function (ELF) unravels the bonding properties of WGeTe on the electronic scale. The values of ELF vary from 0 to 1, representing different types of chemical bonds with various strengths [16]. Both the three-dimensional display and a two-dimensional slice of ELF around W are depicted in . The ELF around W is below 0.68, as shown in (a), which could be reconfirmed by the blue area within the contour line indicating ELF < 0.1, as depicted in (b). Low ELF values indicate that the outer shell electrons around W are delocalized. On the other hand, the electrons are localized around Ge, forming nearly uniform covalent bonds with Te as depicted in (a, b), which is consistent with the bonding environment in pristine GeTe. Besides, as shown in (c), the highest ELF value between W-Te (~ 0.665) lies at 0.846 Å away from Te, which is ~ 1/3 of the W-Te bond length (2.747 Å), demonstrating that the electrons forming W-Te bonds are localized close to Te. Such observation gives another indication that Te accepts outer shell electrons of W, constructing short bonds and resulting in denser structures in WGeTe.

## 4. Conclusions

In summary, by ab initio calculations, we revealed anomalous volume shrinkage and compact local structure when massive W replaces smaller Ge in rock salt GeTe alloys. The electron transfer from W to Te results in shorter ionic W-Te bonds and thus reduced lattice parameters. We demonstrate that electronegativity difference between dopant and host atoms is crucial in determining lattice parameters of complex alloys.

## Acknowledgements

This work is supported by the National Key Research and Development Program of China (Grant No.2017YFB0701700), and the National Natural Science Foundation of China (Grant No.51872017, 51225205).